\documentclass[a4paper,fleqn]{cas-sc}

\usepackage[numbers]{natbib}
\def\tsc#1{\csdef{#1}{\textsc{\lowercase{#1}}\xspace}}
\tsc{WGM}
\tsc{QE}
\tsc{EP}
\tsc{PMS}
\tsc{BEC}
\tsc{DE}
\begin{document}
\let\WriteBookmarks\relax
\def\floatpagepagefraction{1}
\def\textpagefraction{.001}

% Short title
\shorttitle{}

% Short author
\shortauthors{Qing Xu et~al.}

% Main title of the paper
\title [mode = title]{DualAttNet: Synergistic Fusion of Image-level and Fine-Grained Disease Attention for Multi-Label Lesion Detection in Chest X-rays}                      
% Title footnote mark
% eg: \tnotemark[1]
% \tnotemark[1,2]

% Title footnote 1.
% eg: \tnotetext[1]{Title footnote text}
% \tnotetext[<tnote number>]{<tnote text>} 
% \tnotetext[1]{This document is the results of the research
%    project funded by the National Science Foundation.}

% \tnotetext[2]{The second title footnote which is a longer text matter
%    to fill through the whole text width and overflow into
%    another line in the footnotes area of the first page.}

% First author
%
% Options: Use if required
% eg: \author[1,3]{Author Name}[type=editor,
%       style=chinese,
%       auid=000,
%       bioid=1,
%       prefix=Sir,
%       orcid=0000-0000-0000-0000,
%       facebook=<facebook id>,
%       twitter=<twitter id>,
%       linkedin=<linkedin id>,
%       gplus=<gplus id>]
\author{Qing Xu}[style=chinese, orcid=0000-0001-6898-0269]

% Corresponding author indication
\cormark[1]

% Footnote of the first author
% \fnmark[1]

% Email id of the first author
% \ead{cvr_1@tug.org.in}

% URL of the first author
% \ead[url]{www.cvr.cc, cvr@sayahna.org}

%  Credit authorship
% \credit{Conceptualization of this study, Methodology, Software}

% Address/affiliation
\affiliation{organization={The School of Computer Science},
    addressline={University of Lincoln}, 
    city={Lincolnshire},
    % citysep={}, % Uncomment if no comma needed between city and postcode
    postcode={LN6 7TS}, 
    % state={},
    country={United Kingdom}}

% Second author
\author{Wenting Duan}[style=chinese]

% Third author
% \author[2,3]{CV Rajagopal}[%
%    role=Co-ordinator,
%    suffix=Jr,
%    ]
% \fnmark[2]
% \ead{cvr3@sayahna.org}
% \ead[URL]{www.sayahna.org}

% \credit{Data curation, Writing - Original draft preparation}

% Address/affiliation
% \affiliation[2]{organization={Sayahna Foundation},
%     % addressline={}, 
%     city={Jagathy},
%     % citysep={}, % Uncomment if no comma needed between city and postcode
%     postcode={695014}, 
%     state={Trivandrum},
%     country={India}}

% Fourth author
% \author%
% [1,3]
% {Rishi T.}
% \cormark[2]
% \fnmark[1,3]
% \ead{rishi@stmdocs.in}
% \ead[URL]{www.stmdocs.in}

% \affiliation[3]{organization={STM Document Engineering Pvt Ltd.},
%     addressline={Mepukada}, 
%     city={Malayinkil},
%     % citysep={}, % Uncomment if no comma needed between city and postcode
%     postcode={695571}, 
%     state={Trivandrum},
%     country={India}}

% % Corresponding author text
% \cortext[cor1]{Corresponding author}
% \cortext[cor2]{Principal corresponding author}

% % Footnote text
% \fntext[fn1]{This is the first author footnote. but is common to third
%   author as well.}
% \fntext[fn2]{Another author footnote, this is a very long footnote and
%   it should be a really long footnote. But this footnote is not yet
%   sufficiently long enough to make two lines of footnote text.}

% % For a title note without a number/mark
% \nonumnote{This note has no numbers. In this work we demonstrate $a_b$
%   the formation Y\_1 of a new type of polariton on the interface
%   between a cuprous oxide slab and a polystyrene micro-sphere placed
%   on the slab.
%   }

% Here goes the abstract
\begin{abstract}
Chest radiographs are the most commonly performed radiological examinations for lesion detection. Recent advances in deep learning have led to encouraging results in various thoracic disease detection tasks. Particularly, the architecture with feature pyramid network performs the ability to recognise targets with different sizes. However, such networks are difficult to focus on lesion regions in chest X-rays due to their high resemblance in vision. In this paper, we propose a dual attention supervised module for multi-label lesion detection in chest radiographs, named DualAttNet. It efficiently fuses global and local lesion classification information based on an image-level attention block and a fine-grained disease attention algorithm. A binary cross entropy loss function is used to calculate the difference between the attention map and ground truth at image level. The generated gradient flow is leveraged to refine pyramid representations and highlight lesion-related features. We evaluate the proposed model on VinDr-CXR, ChestX-ray8 and COVID-19 datasets. The experimental results show that DualAttNet surpasses baselines by 0.6\% to 2.7\% mAP and 1.4\% to 4.7\% AP$_{50}$ with different detection architectures. The code for our work and more technical details can be found at \url{https://github.com/xq141839/DualAttNet}.
\end{abstract}

% Use if graphical abstract is present
% \begin{graphicalabstract}
% \includegraphics{figs/grabs.pdf}
% \end{graphicalabstract}

% Research highlights
% \begin{highlights}
% \item Research highlights item 1
% \item Research highlights item 2
% \item Research highlights item 3
% \end{highlights}

% Keywords
% Each keyword is seperated by \sep

\maketitle

\section{Introduction}

Chest X-ray (CXR), as a cornerstone of X-ray exam, has become the most frequent radiological screening since the last century. Due to cost efficiency and low radiational doses, CXR is usually an essential approach for diagnosing different types of thoracic diseases \cite{raoof2012interpretation}. Based on the location and direction of the patient with respect to the X-ray generator and detector, CXR can be separated into three main categories: posteroanterior, anteroposterior, and lateral. Where posteroanterior and anteroposterior images are mainly used to identify the position of the disease. Some of the ambiguous lesions are recalibrated using lateral images \cite{lai2012diagnostic}. In addition, with the development of computing devices, constructing models used for automatic CXR analysis is no longer impossible. Various algorithms have been introduced, which can optimise monotonous tasks, improve the sensitivity for rare cases and assist with long-range diagnosis \cite{yan2012automatic, solti2009automated, yu2011automatic}.

Multi-label abnormality detection plays a crucial role in thoracic disease diagnosis. Due to the high visual similarity of CXRs and the complexity of interpretation, it can be also known as the fine-grained image recognition task. Existing models, combined with Deep Neural Networks (DNNs) and various learning strategies, have made a significant breakthrough in automatic localisation of thoracic diseases \cite{rajpurkar2017chexnet, girshick2015fast, bochkovskiy2020yolov4}. The architecture of these methods mainly combines deep-layer encoders \cite{he2016deep, huang2017densely} and Feature Pyramid Network (FPN) \cite{lin2017feature} to extract features from CXRs. The main idea of FPN is to adopt multi-scale fusion to improve the performance of identifying targets with different sizes. The final layers of these networks involve two sub-networks consisting of a header for regressing the bounding box and predicting classification. These proposed algorithms are also called one-stage detection networks, which usually perform faster inference but lower precision compared to the two-stage models. To improve the performance of the one-stage models, embedding an attention mechanism module is a common solution. Attention algorithms work in a way to guide networks to focus on lesion zones by applying different non-linear transformations and combinations on the feature map. However, the additional computation costs reduce the inference speed of the network and practical applicability. Also, CXR images may include different disease types. Existing methods lack the supervision of classification and have the risk of false positive detections, which can be a serious issue for practical diagnostic applications.

To address this issue, we propose a dual attention module for multi-label lesion detection in chest X-rays, called DualAttNet, involving two components. The multi-scale feature maps extracted from FPN will be first fed into an Image-level Attention (ILA) block that adaptively recalibrates the weight of each channel. In this case, ILA is able to capture global classification information from the whole feature map but lacks supervision for the location of different disease types. For this reason, we design another fine-grained disease attention (FGDA) algorithm to guide the network to pay attention to the area of interest. It is connected to the header of the detector, utilises predicted anchors to enhance spatial representations, and combines with global attention features from ILA. Finally, the fusion feature map, including classification and location information, will be compared to the lesion-level one-hot labels using binary cross entropy loss function so that the backward gradients can help suppress feature activation in irrelevant regions. The main contributions of this work can be summarised as follows:
\begin{enumerate}[1)]

    \item An ILA model is presented to receive global classification information from the features in FPN layers. Additionally, we introduce an FGDA algorithm to represent local attention information of features from the header of the detection architecture.
    \item We use FGDA to improve the representations from ILA and integrate a new DualAttNet. The outputs will compare with the ground truth from the image level. As the proposed model is not in charge of generating refined feature maps, embedding the module will not decrease the inference speed of the detector.
    \item Our experiments are conducted with three different CXR datasets: VinDr-CXR \cite{nguyen2022vindr}, ChestX-ray8 \cite{wang2017chestx}, and COVID-19 \cite{cohen2020covid}. Evaluation results demonstrate that our proposed DualAttNet performs better than other attention models in terms of standard detection metrics - average precision (AP). It can be a new state-of-the-art (SOTA) method for multi-label lesion detection in CXRs.
\end{enumerate}

\section{Related Work}
\subsection{Feature Pyramid Network for Object Detection}
Feature Pyramid Network (FPN) \cite{lin2017feature} has been widely used in one-stage and two-stage object detection algorithms. To fully utilise the semantic information from each down-sampling layer, it constructs a top-down architecture with horizontal connections, which combines the feature maps at different scales. Many studies have demonstrated that FPN can better recognise objects with various sizes \cite{zou2023object}. Existing detectors usually leverage FPN to fuse multi-scale features. RetinaNet \cite{lin2017focal}, proposed by Lin et al., simply uses multi-layer Convolutional Neural Networks (CNNs) to transform every pyramid feature map to potential bounding boxes and corresponding classes of targets. To address the issue of imbalanced datasets, RetinaNet is trained with the focal loss function that adjusts the weight between easy and hard samples. Redmon et al. \cite{redmon2016you} presented a one-stage object detection model, called You Only Look Once (YOLO). The model first divides the input image into an S $\times$ S (e.g., 7 $\times$ 7) grid. The grid cell containing the centre of the labeled bounding box is responsible for detecting the object. The experimental results show that YOLO surpasses the baseline and achieves faster inference. However, YOLO has a limited ability to detect adjacent objects because only one object can be recognised by each grid. To improve the performance of YOLO, various updated versions, replacing the backbone or header and embedding FPN, have been published \cite{chen2021you, ge2021yolox, li2022yolov6}. Furthermore, Zhang et al. \cite{zhang2021varifocalnet} dedicated to the improvement of object localisation and introduced a VFNet for dense object detection. The model corrects the bounding box feature representation from FPN and is trained with a new varifocal loss function, which refines the predicted bounding boxes using an IoU-aware classification score. Extensive experiments demonstrate the VFNet model outperforms previous SOTA methods. However, these architectures mainly focus on the optimisation of localising targets. In contrast, our method improves the sensitivity for thoracic disease detection. In other words, it suppresses the detection of irrelevant disease regions and generates a low false positive rate, which is beneficial for clinical applications.

\subsection{Attention Mechanism}
In the last few years, the attention mechanism has achieved significant success in computer vision. It aims to emphasise target regions while suppressing irrelevant features in an image. Hu et al. \cite{hu2018squeeze} proposed a Squeeze-and-Excitation Network (SENet) that captures channel-wise information and re-weights feature representations. This module can be flexibly embedded into any feature-extract layer to guide what networks should pay attention to. Following SENet, different variants \cite{gao2019global, wang2020eca, qin2021fcanet} have been designed to further handle channel features efficiently. Spatial attention is another mechanism used for adaptive spatial area selection. Especially, Vision Transformer (ViT), proposed by Dosovitskiy et al. \cite{dosovitskiy2020image}, has received the most attention recently. Multi-head attention module is the main idea of ViT, which calculates dynamic spatial-temporal correlations from an input sequence. Inspired by ViT, various transformer-based frameworks \cite{yuan2021tokens, liu2021swin, he2022masked} have been constructed and show better performance than CNNs. However, such models usually include a large number of parameters and require a high cost to configure a trainable environment. In addition, Woo et al., \cite{woo2018cbam} presented the convolutional block attention module (CBAM) that combines two different dimensional information: channel and spatial. It activates valuable channels as well as highlights informative local regions. Existing attention modules receive features from the encoder and directly produce optimised feature maps, which are integrated into the whole model. On the contrary, our approach leverage both global and local gradient flows to refine the feature representation from FPN.

\section{Methodology}
In this section, we describe the components of the proposed DualAttNet. They connect with each feature pyramid layer and classification head of the detection architecture. The outputs integrate global and lesion-level local attention gradient flows that are used for guiding the detector to focus the disease regions in chest X-rays.

\subsection{Image-level Attention Block}
\begin{figure}[!thbp]
  \centering
  % \includesvg[inkscapelatex=false]{dsh.svg}
  \includegraphics[width=1\linewidth]{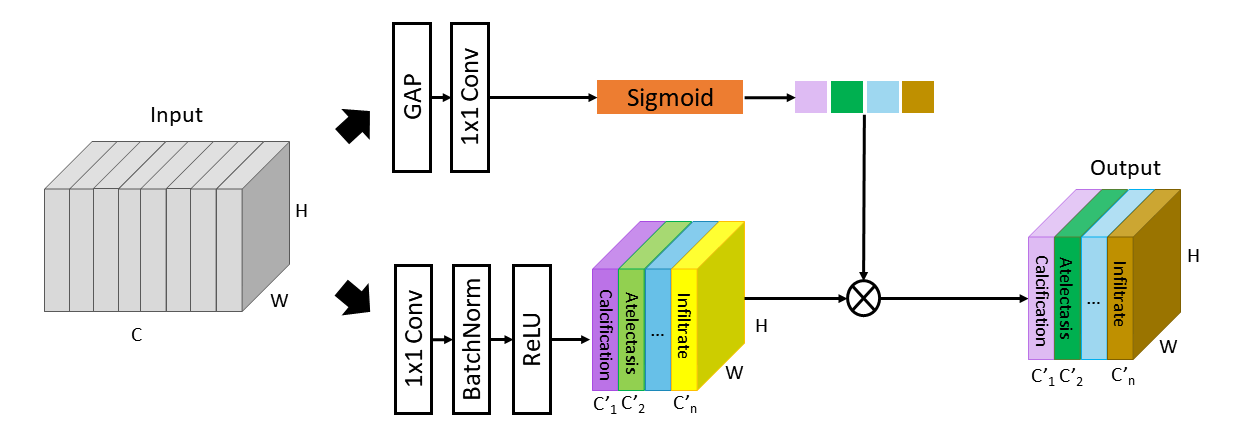}
  \caption{The structure of image-level attention block.}
  \label{fig:f1}
\end{figure}

For the detection architecture in CXRs, the feature map of each layer in FPN contains thoracic disease objects at different scales. To distinguish various lesion types efficiently, we propose an image-level attention (ILA) block. The architecture of ILA is provided in Figure \ref{fig:f1}.
Firstly, a multi-scale feature map $X_k$ obtained by FPN is considered as input. It will go through two different branches. On one branch (i.e. lower branch shown in Figure \ref{fig:f1}), we use 1 $\times$ 1 convolution, followed by a BatchNorm and ReLU activation, to compress the channel-wise information, which can be represented as:
\begin{equation}
F_k=ReLU(BN(Conv(X_k)))
\end{equation}
Where the channel of the output feature $F_k$ is equal to the number of disease classes and each channel stands for one type of lesion. On the other branch, a global average pooling (GAP) is used to capture global spatial information from the feature map $X_k$. The c-th element of the pooled feature $S_k$ is calculated as:
\begin{equation}
S^k_c=\frac{1}{H\times W}\sum_{i=1}^{H}\sum_{j=1}^{W}{{X}^k_c\left(i,j\right)}
\end{equation}
The GAP compresses spatial dimensions and generates channel-wise statistics. We also connect 1x1 convolution with GAP for channel alignment with $F_k$. A soft attention is used across channels to adaptively select different spatial scales, which is guided by $S_k$. A soft assignment weight is designed by:
\begin{equation}
Att_k = \sigma(S^k_c)=\frac{1}{1 + e^{(-S^k_c)}}
\end{equation}
Here we apply the sigmoid to re-calibrate the weight of the feature channel (i.e. different lesions). For the softmax function, it aims to improve the probability of the correct category while suppressing the remaining categories. Instead, there may be more than one lesion in CXRs. Therefore, we do not consider the softmax function in our module. Finally, an operation of element multiplication $\bigodot$ is performed on the refined weight $Att_k$ and the corresponding feature map $F_k$:
\begin{equation}
Y'_k = S^k \odot Att_k, \ \ k \in 1, 2, 3 \cdots N
\end{equation}
Where k is the number of feature maps extracted from FPN. In sum, the ILA module aggregates global contextual information for different thoracic diseases.

\subsection{Fine-Grained Disease Attention}
\begin{figure}[!thbp]
  \centering
  % \includesvg[inkscapelatex=false]{dsh.svg}
  \includegraphics[width=0.6\linewidth]{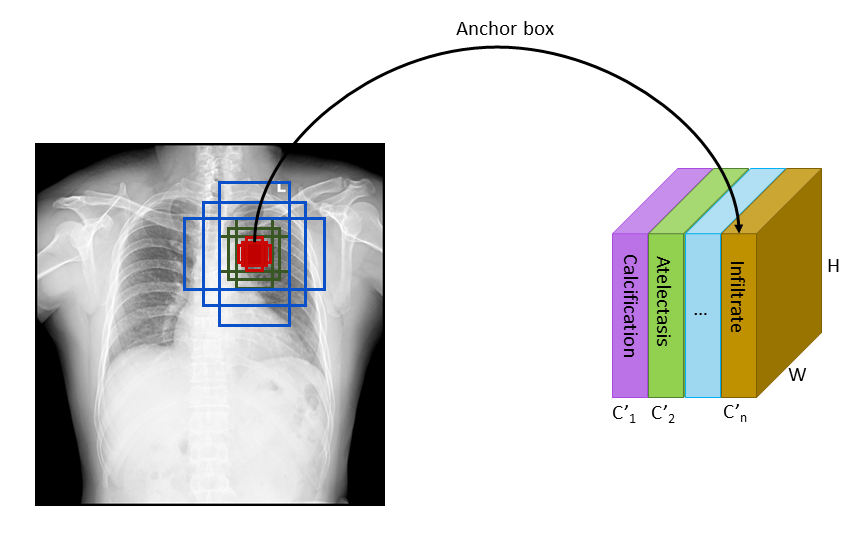}
  \caption{The illustration of fine-grained disease attention algorithm.}
  \label{fig:f2}
\end{figure}

Anchor-based networks \cite{zhang2018single, redmon2018yolov3} have achieved great success in object detection. Anchors can be known as a set of predefined bounding boxes with different sizes and aspect ratios. To select a suitable anchor for detecting objects, we compute the intersection over union (IoU) between each anchor and targets. The head of detection architectures usually adopts small fully convolutional networks to classify and regress the number of bounding boxes based on the anchor configuration. As the output feature map involves position and class information of disease, it can be transformed into a local attention map. Therefore, we construct a fine-grained disease attention (FGDA) algorithm, which is shown in Figure \ref{fig:f2}.
For each disease type $n$, it first extracts the anchor information (Height $\times$ Width $\times$ Anchor) from the classification feature map $X’_k$. Where the anchor with the maximum prediction is used as the attention map, which can be represented as:
\begin{equation}
A = \arg \max (X'_k) ,\ \ A\in{R}^{H\times W\times Anchor}
\end{equation}
To convert the prediction values to probabilities, a simple normalisation method is performed:
\begin{equation}
A^n_{i,j} = \frac{{A}^n_{i,j}}{\sum_{\alpha=1}^H\sum_{\beta=1}^W A^n_{\alpha,\beta}}
\end{equation}
Finally, we concatenate the attention map of all classes:
\begin{equation}
Y''_k = Cat([A^1, A^2, \cdots A^n])
\end{equation}
As illustrated by the above analysis, our proposed FGDA module can extract the location information of different lesions from CXRs.

\subsection{DualAttNet Architecture}
\begin{figure}[!thbp]
  \centering
  % \includesvg[inkscapelatex=false]{dsh.svg}
  \includegraphics[width=1\linewidth]{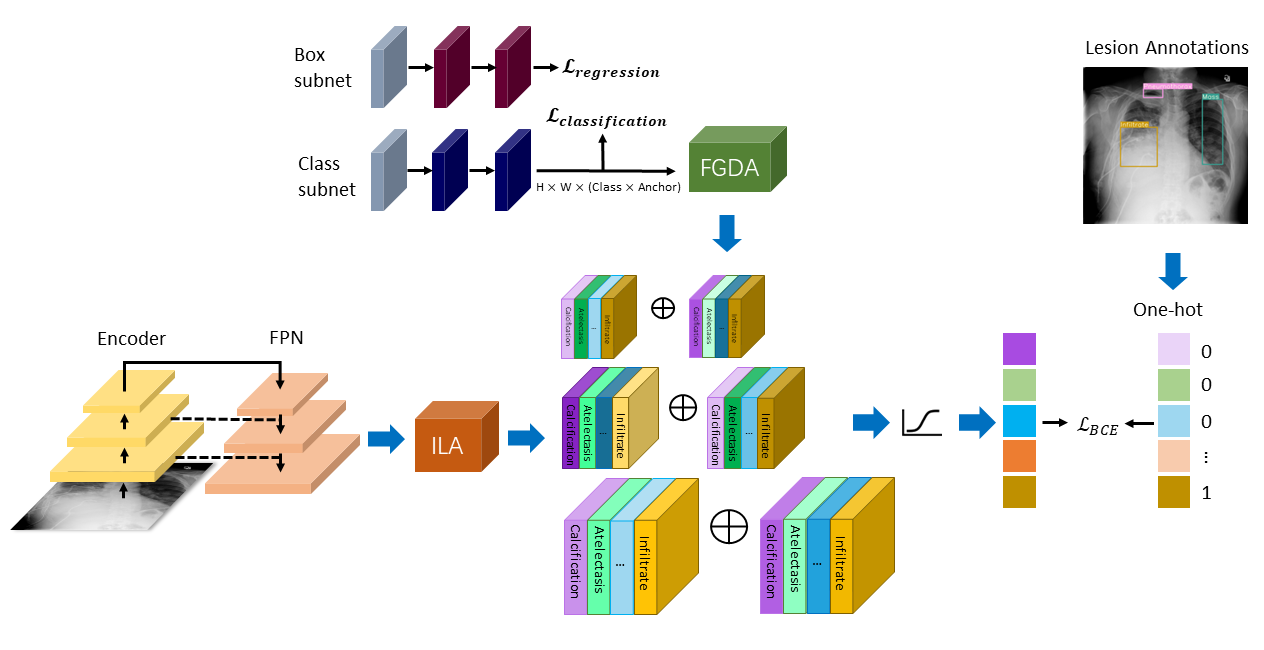}
  \caption{The overview of our DualAttNet architecture connected with RetinaNet \cite{lin2017focal}.}
  \label{fig:f3}
\end{figure}

Modern object detection architectures often focus on the enhancement of location precision. For lesion recognition in CXRs, false positive detection is a major issue in clinical applications. In this case, we embed the newly designed attention module into a so-called DualAttNet to guide the network focusing on the symptomatic areas while suppressing unrelated detection. An overview of the proposed architecture is presented in Figure \ref{fig:f3}.
The input image first goes through an encoder and FPN. The feature map of the FPN layers will first be fed to the ILA module to extract global attention information. Then local attention is obtained from the FGDA module. These attention maps provide two different regions of interests (ROIs). Then, we combine the global and local attention information and reshape the integrated map to a classification vector, which can be computed as:
\begin{equation}
\hat{Y}=\sigma\left(\frac{1}{K}\sum_{k=1}^{K}\sum_{i=1}^{H}\sum_{j=1}^{W}{{Y}'_k\left(i,j\right)+{Y}''_k\left(i,j\right)}\right),\ \ \hat{Y}\in{R}^{Batch\times Class}
\end{equation}
Where $\sigma$ is the sigmoid function and $k$ is the index of FPN layers. This operation also combines multi-scale feature maps in each pyramid layer, which can perceive different sizes of lesions in medical imaging. To refine the inclusion of multiple symptomatic lesions being represented in the pyramid feature, we adopt the binary cross entropy loss as the objective function. It can be defined as:
\begin{equation}
\mathcal{L}_{\mathrm{BCE}}(\hat{Y}, Y)=-\sum_{n=1}^{N}\left[Y_{n} \cdot \log \left(\hat{Y}_{n}\right)+\left(1-Y_{n}\right) \cdot \log \left(1-\hat{Y}_{n}\right)\right]
\end{equation}
Where $Y_{n}$ is the lesion-level annotations and $\hat{Y}_{n}$ is the prediction. As a result, the gradient from both global and local branches can optimise the feature representation of FPN for lesion localisation and classification in CXRs. The proposed DualAttNet can adapt to any FPN and classification sub-network without affecting the inference speed of the original detection architecture.

\section{Experiments and Results}
\subsection{Dataset}
\label{sec:s1}
To demonstrate the effectiveness of DualAttNet, we evaluate it on three public datasets of multi-label thoracic disease detection:
\begin{itemize}

    \item VinDr-CXR \cite{nguyen2022vindr} comprises 4394 Digital Imaging and Communications in Medicine (DICOM) files with 14 types of thoracic abnormalities from chest radiographs. It is also the training database for the Kaggle 2021 VinBigData Chest X-ray Abnormalities Detection Competition.
    \item The second dataset used in this study is ChestX-ray8 \cite{wang2017chestx}. It includes 108,948 posteroanterior X-ray images of 32,717 unique patients with the eight disease labels. Particularly, 880 images have annotated objects with bounding boxes, which can be used as the ground truth to evaluate the disease localization performance.
    \item In order to evaluate the robustness of the proposed architecture on a small dataset, we add a COVID-19 dataset \cite{cohen2020covid} to our experiment. There are only 223 frontal X-ray images for the detection of five pneumonia types: COVID-19, SARS, Streptococcus, Pneumocystis, and ARDS.
\end{itemize}

A fixed random seed is used to divide all datasets into three sets: train, validation, and test, in the ratio of 7:2:1. More details about the data split are provided in Table \ref{tab:t1}.
\begin{table}[!h]
  \caption{Details of the chest-Xray datasets used in our experiments.}\label{tab:t1}
  {\scalebox{1}{
  \begin{tabular}{llllll}
  \toprule
  Dataset &  Images &  Input size &  Train & Valid & Test\\
  \midrule
  VinDr-CXR & 4394 & Variable & 3074 & 880 & 440\\
  ChestX-ray8 & 880 & 1024x1024 & 616 & 176 & 88\\
  COVID-19 & 223 & Variable & 154 & 46 & 23\\
  \bottomrule
  \end{tabular}}}
\end{table}

\subsection{Evaluation Metrics}
In the field of object detection, average precision (AP) is the standard evaluation metric. Specifically, we calculate the mean AP (mAP) from AP$_{50}$ to AP$_{95}$ with an interval of 5, following the COCO evaluator. Recall plays an important role in clinical applications \cite{anwar2018medical}. Therefore, we also compute average recall (AR) for small (AR$_S$), medium (AR$_M$), and large (AR$_L$) lesions. They are respectively defined as being between $0^2$ to $32^2$, $32^2$ to $96^2$, and $96^2$ to $1e5^2$ pixels in area. All reported statistics were averaged over the number of abnormalities in the dataset.

\subsection{Data Augmentation}
As illustrated in the section \ref{sec:s1}, medical datasets contain a limited number of samples. The model is easily prone to overfitting in the training stage. To mitigate this issue, we adopt data augmentation approaches to enrich the diversity of images and improve the robustness of the model. In the experiment, horizontal flip, rotation, brightness, contrast, and cutout transformations are randomly applied to the training set of each dataset with a probability of 0.1.

\subsection{Implementation Details}
All experiments are implemented using PyTorch 1.10.0 framework on a single NVIDIA RTX 2080Ti GPU, 4-core CPU, and 28GB RAM. In order to compare DualAttNet with other attention modules, we use a commonly FPN-based detector, RetinaNet \cite{lin2017focal} as the fundamental architecture, a pretrained ResNet-50 \cite{he2016deep} as the encoder, focal loss as the main objective function and Adam as the optimiser with a learning rate of 1e-5. The number of batch sizes and epochs are set to 4 and 40 respectively. In the training phase, the images from all three datasets are resized to 512 $\times$ 512. We also apply ReduceLROnPlateau to adjust the learning rate, where the hyperparameters: patience and factor are set to 3 and 0.1 respectively. All experiments on three datasets are conducted on the same training, validation, and testing datasets. 

\subsection{Results}
In this section, we show quantitative results on three different CXR datasets and compare our proposed model with other SOTA methods.

\subsubsection{Comparison on VinDr-CXR Dataset}
\begin{table}[h]
  \centering
  \small
  \setlength\tabcolsep{5pt}
  \caption{Results on the VinDr-CXR dataset.}
  {\scalebox{1}{
  \begin{tabular}{llllllllll}
  \toprule
  Methods & mAP & AP$_{50}$ & AP$_{75}$ & AP$_{S}$ & AP$_{M}$ & AP$_{L}$ & AR$_{S}$ & AR$_{M}$ & AR$_{L}$\\
  \midrule
  SENet \cite{hu2018squeeze} & 0.103 & 0.219 & 0.104 & 0.004 & 0.085 & 0.114 & 0.023 & 0.168 & 0.247 \\
  CBAM \cite{woo2018cbam} & 0.098 & 0.214 & 0.096 & \textbf{0.010} & 0.087 & 0.092 & \textbf{0.032} & 0.170 & 0.225 \\
  PSA \cite{zhang2022epsanet} & 0.103 & 0.221 & 0.099 & 0.005 & 0.086 & 0.116 & 0.025 & 0.167 & 0.230 \\
  CCNet \cite{huang2019ccnet} & 0.101 & 0.217 & 0.102 & 0.002 & 0.084 & 0.107 & 0.025 & 0.169 & 0.242 \\
  ACmix \cite{pan2022integration} & 0.099 & 0.215 & 0.099 & 0.005 & 0.082 & 0.107 & 0.029 & 0.171 & 0.236 \\
  CoT \cite{li2022contextual} & 0.104 & 0.218 & 0.102 & 0.003 & 0.090 & 0.112 & 0.025 & 0.167 & 0.240 \\
  ParNet \cite{goyal2022non} & 0.102 & 0.221 & 0.105 & 0.006 & 0.085 & 0.117 & 0.024 & 0.163 & 0.233 \\
  DualAttNet (Ours) & \textbf{0.116} & \textbf{0.241} & \textbf{0.114} & 0.005 & \textbf{0.098} & \textbf{0.121} & 0.027 & \textbf{0.177} & \textbf{0.257}\\ 
  \bottomrule
  \end{tabular}}}
  \label{tab:t2}
\end{table}
The quantitative result on VinDr-CXR dataset is presented in Table \ref{tab:t2}. For lesion detection in chest X-rays, a series of AP metrics are able to reveal the precision of the location. From the Table, DualAttNet shows better performance than other SOTA models in most of the metrics. Specifically, our module achieves an mAP of 0.116 and an AP$_{50}$ of 0.241, which outperforms PSA by 1.3\% in terms of mAP and 2.0\% in AP$_{50}$. Also, recall is an essential metric in clinical applications. In the experimental result, DualAttNet also displays higher AR$_M$ and AR$_L$ scores than other methods.

\subsubsection{Comparison on ChestX-ray8 Dataset}
\begin{table}[h]
  \centering
  \small
  \setlength\tabcolsep{5pt}
  \caption{Results on the ChestX-ray8 dataset.}
  {\scalebox{1}{
  \begin{tabular}{llllcllcll}
  \toprule
  Methods & mAP & AP$_{50}$ & AP$_{75}$ & AP$_{S}$ & AP$_{M}$ & AP$_{L}$ & AR$_{S}$ & AR$_{M}$ & AR$_{L}$\\
  \midrule
  SENet \cite{hu2018squeeze} & 0.054 & 0.124 & 0.036 & -- & 0.002 & 0.065 & -- & 0.034 & 0.140 \\
  CBAM \cite{woo2018cbam} & 0.057 & 0.126 & 0.050 & -- & 0.008 & 0.074 & -- & 0.057 & 0.142 \\
  PSA \cite{zhang2022epsanet} & 0.061 & 0.123 & 0.050 & -- & 0.019 & 0.076 & -- & 0.043 & 0.152 \\
  CCNet \cite{huang2019ccnet} & 0.063 & 0.132 & 0.077 & -- & 0.006 & 0.074 & -- & 0.039 & 0.154 \\
  ACmix \cite{pan2022integration} & 0.053 & 0.104 & 0.063 & -- & 0.008 & 0.061 & -- & \textbf{0.069} & 0.142 \\
  CoT \cite{li2022contextual} & 0.062 & 0.133 & 0.042 & -- & 0.016 & 0.075 & -- & 0.028 & 0.160 \\
  ParNet \cite{goyal2022non} & 0.061 & 0.108 & \textbf{0.078} & -- & 0.001 & 0.073 & -- & 0.020 & 0.146 \\
  DualAttNet (Ours) & \textbf{0.071} & \textbf{0.145} & 0.076 & -- & \textbf{0.026} & \textbf{0.076} & -- & 0.056 & \textbf{0.161}\\ 
  \bottomrule
  \end{tabular}}}
  \label{tab:t3}
\end{table}
The annotation in medical imaging is usually expensive and time-consuming. Therefore, many CXR datasets contain limited box annotations, which is a challenge for generalisation of the model. For lesion detection in chest X-rays, the performance of the architecture on mAP and AP$_{50}$ are the most witnessed metrics. A comparison between each model is provided in Table \ref{tab:t3}. We are not able to calculate AP$_S$ and AR$_S$ metrics because there are no small targets based on COCO criteria.  The result demonstrates that DualAttNet has an increase of 0.8\% over CCNet in mAP and 1.2\% in AP$_{50}$. Particularly, our proposed model presents a significant enhancement over the recent self-attention architecture, where the mAP of DualAttNet is 2.8\% higher than ACmix, and the AP$_{50}$ of DualAttNet is 4.1\% higher than this model.

\subsubsection{Comparison on COVID-19 Dataset}
\begin{table}[h]
  \centering
  \small
  \setlength\tabcolsep{5pt}
  \caption{Results on the COVID-19 dataset.}
  {\scalebox{1}{
  \begin{tabular}{llllcclccl}
  \toprule
  Methods & mAP & AP$_{50}$ & AP$_{75}$ & AP$_{S}$ & AP$_{M}$ & AP$_{L}$ & AR$_{S}$ & AR$_{M}$ & AR$_{L}$\\
  \midrule
  SENet \cite{hu2018squeeze} & 0.015 & 0.049 & 0.010 & -- & -- & 0.015 & -- & -- & 0.040 \\
  CBAM \cite{woo2018cbam} & 0.017 & 0.041 & 0.005 & -- & -- & 0.017 & -- & -- & 0.042 \\
  PSA \cite{zhang2022epsanet} & 0.014 & 0.041 & 0.003 & -- & -- & 0.014 & -- & -- & 0.036 \\
  CCNet \cite{huang2019ccnet} & 0.013 & 0.050 & 0.000 & -- & -- & 0.013 & -- & -- & 0.037 \\
  ACmix \cite{pan2022integration} & 0.013 & 0.032 & 0.002 & -- & -- & 0.013 & -- & -- & 0.045 \\
  CoT \cite{li2022contextual} & 0.016 & 0.045 & 0.003 & -- & -- & 0.016 & -- & -- & 0.037 \\
  ParNet \cite{goyal2022non} & 0.016 & 0.051 & \textbf{0.014} & -- & -- & 0.018 & -- & -- & 0.047 \\
  DualAttNet (Ours) & \textbf{0.021} & \textbf{0.064} & 0.010 & -- & -- & \textbf{0.019} & -- & -- & \textbf{0.053}\\ 
  \bottomrule
  \end{tabular}}}
  \label{tab:t4}
\end{table}
COVID-19 has tremendously impacted patients and the medical system globally. Current deep learning-based studies are expected to distinguish COVID-19 from other types of pneumonia automatically. To meet the requirement, we evaluate all models on the COVID-19 dataset. As existing annotations are based on the whole area of the lung, the dataset does not include any small or medium objects, and related metrics are not computed in the experiment. The quantitative results are provided in Table \ref{tab:t4}. We can observe that DualAttNet performs an mAP of 0.021 with a rise of 0.4\% over CBAM and 1.3\% in AP$_{50}$ compared to the ParNet model.

\subsection{Ablation Study}
\begin{table}[!bthp]
  \caption{Detailed ablation study of the DualAttNet architecture.}
  {\scalebox{1}{
  \begin{tabular}{lllllccccccl}
  \toprule
  Dataset & Method & mAP & AP$_{50}$ & AP$_{75}$ & AP$_{S}$ & AP$_{M}$ & AP$_{L}$ & AR$_{S}$ & AR$_{M}$ & AR$_{L}$\\
  %& Size & & & & & & & & &
  \midrule
  & RetinaNet \cite{hu2018squeeze} & 0.104 & 0.219 & 0.097 & 0.003 & 0.091 & 0.119 & 0.022 & 0.155 & 0.219\\
  \multirowcell{2}{VinDr-CXR} & RetinaNet + ILAB & 0.111 & 0.228 & 0.103 & \textbf{0.007} & 0.093 & 0.122 & 0.030 & 0.175 & 0.242\\
  & RetinaNet + FGDA & 0.115 & 0.232 & 0.108 & 0.006 & 0.092 & \textbf{0.124} & \textbf{0.032} & 0.162 & 0.230\\
  & RetinaNet + DualAttNet & \textbf{0.116} & \textbf{0.241} & \textbf{0.114} & 0.005 & \textbf{0.098} & 0.121 & 0.027 & \textbf{0.177} & \textbf{0.257}\\
  \midrule
  & RetinaNet \cite{hu2018squeeze} & 0.049 & 0.106 & 0.033 & -- & 0.004 & 0.061 & -- & 0.059 & 0.145\\
  \multirowcell{2}{ChestX-ray8} & RetinaNet + ILAB & 0.066 & 0.124 & 0.056 & -- & 0.013 & 0.072 & -- & \textbf{0.066} & 0.160\\
  & RetinaNet + FGDA & 0.062 & 0.119 & 0.061 & -- & 0.006 & 0.074 & -- & 0.024 & 0.154\\
  & RetinaNet + DualAttNet & \textbf{0.071} & \textbf{0.145} & \textbf{0.076} & -- & \textbf{0.026} & \textbf{0.076} & -- & 0.056 & \textbf{0.161}\\  
  \midrule
  & RetinaNet \cite{hu2018squeeze} & 0.011 & 0.037 & 0.000 & -- & -- & 0.011 & -- & -- & 0.037\\
  \multirowcell{2}{COVID-19} & RetinaNet + ILAB & 0.018 & 0.053 & 0.007 & -- & -- & 0.016 & -- & -- & 0.038\\
  & RetinaNet + FGDA & 0.015 & 0.045 & 0.009 & -- & -- & 0.013 & -- & -- & 0.042\\
  & RetinaNet + DualAttNet & \textbf{0.019} & \textbf{0.064} & \textbf{0.010} & -- & -- & \textbf{0.019} & -- & -- & \textbf{0.053}\\
  
  \bottomrule
  \end{tabular}}}
  \label{tab:t5}
\end{table}
\subsubsection{Efficiency of ILA Block}
The DualAttNet model adopts the ILA block to capture global spatial information of different thoracic diseases at image level, which can refine multi-scale feature maps from FPN layers. The effectiveness of ILA block can be evaluated by comparing the configurations: RetinaNet and RetinaNet + ILA in Table \ref{tab:t5}. In terms of the mAP and AP$_{50}$, the ILA block respectively produces an improvement of 0.7\% and 0.9\% on the VinDr-CXR dataset, 1.7\% and 1.8\% increase on the ChestX-ray8 dataset, as well as 0.7\% and 1.6\% enhancement on the COVID-19 dataset. Thus, it can be concluded that the ILA block enhances the performance of the original RetinaNet.

\subsubsection{Significance of FGDA algorithm}
The FGDA algorithm is an essential part of the proposed DualAttNet model. It extracts anchor information of each lesion from the classification sub-network and then transformed into local attention feature maps uses with a normalisation method. We compare the network configurations: RetinaNet and RetinaNet + FGDA to evaluate the effectiveness of the FGDA algorithm. From the mAP and AP$_{50}$ metrics in Table \ref{tab:t5}, FGDA respectively shows an improvement of 1.1\% and 1.3\% on the VinDr-CXR dataset, 1.3\% and 1.3\% improvement on the ChestX-ray8 dataset, as well as 0.4\% and 0.8\% improvement on the COVID-19 dataset. We can argue that the FGDA-embedded architecture performs better than the RetinaNet model. Overall, ILA has a more significant impact than the FGDA algorithm. By taking advantage of both modules, the AligNet model (RetinaNet + ILA + FGDA) can further improve the mAP by 0.1\% to 0.9\% and the AP$_{50}$ by 0.9\% to 2.6\% compared to the RetinaNet with a single ILA or FGDA module.

\subsection{Applicability of DualAttNet}
\begin{figure}[!thbp]
  \centering
  % \includesvg[inkscapelatex=false]{dsh.svg}
  \includegraphics[width=1\linewidth]{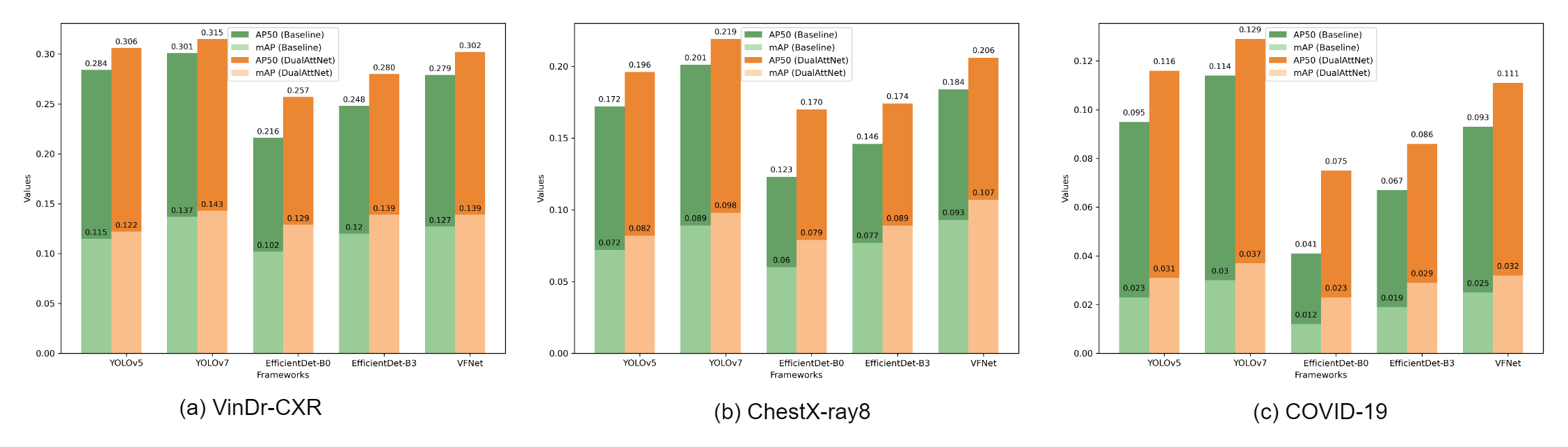}
  \caption{The performance comparison of baselines vs DualAttNet-embedded detection architectures on three CXR datasets.}
  \label{fig:f4}
\end{figure}

We also transfer DualAttNet to other popular detection frameworks. In detail, the ILA and FGDA modules are connected with their feature pyramid layers and classification header respectively. Figure \ref{fig:f4} displays the comparison results between original and DualAttNet-based detectors, including YOLOv5 \cite{jocher2022ultralytics}, YOLOv7 \cite{wang2023yolov7}, EfficientDet-B0 \cite{tan2020efficientdet}, EfficientDet-B3 \cite{tan2020efficientdet} and VFNet \cite{zhang2021varifocalnet}. It can be demonstrated that adding DualAttNet can further boost the mean mAP by around 1.2\% and mean AP$_{50}$ by around 2.5\% on three CXR datasets.

\section{Discussion}
\begin{figure}[!thbp]
  \centering
  % \includesvg[inkscapelatex=false]{dsh.svg}
  \includegraphics[width=0.6\linewidth]{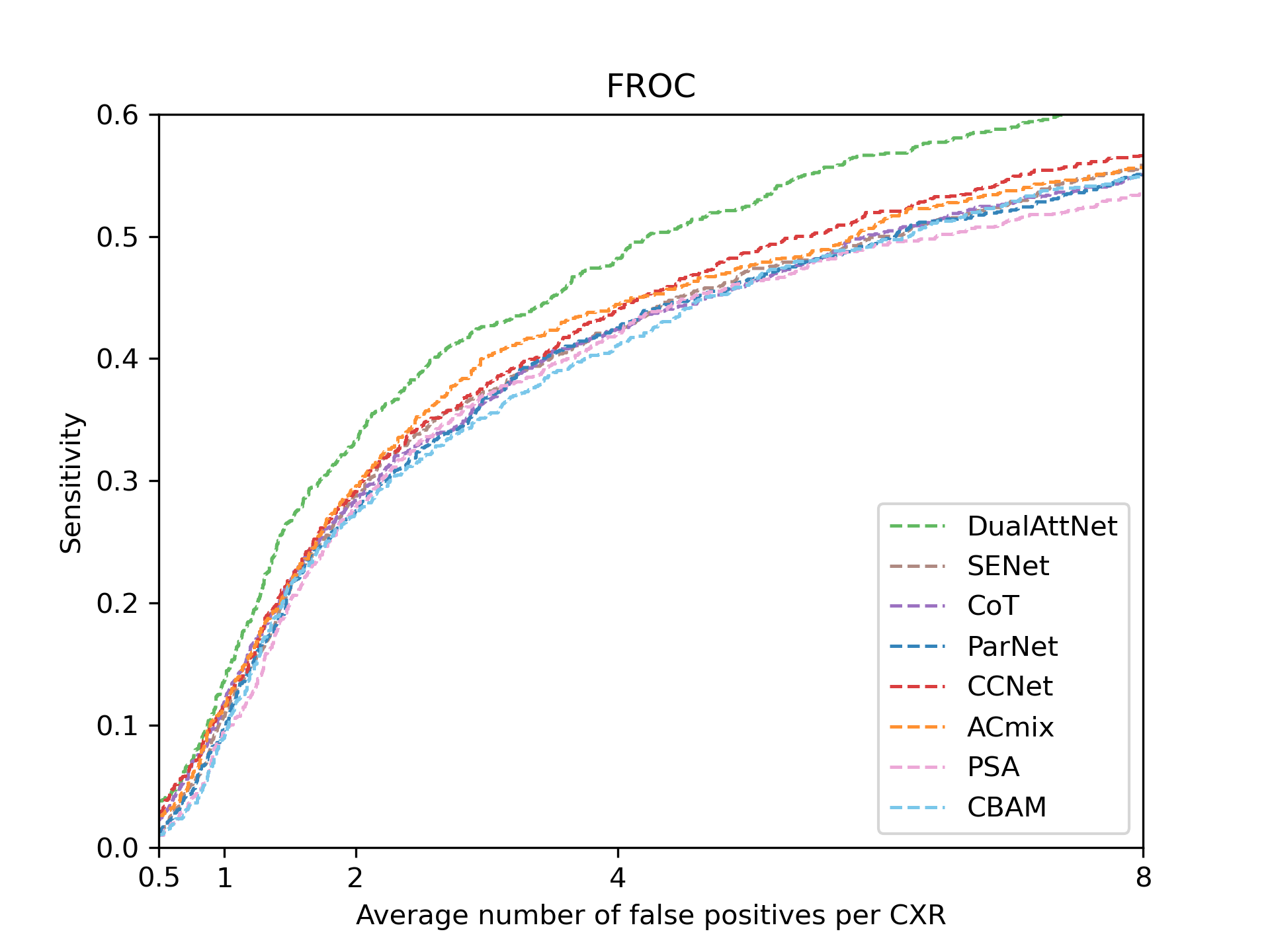}
  \caption{The comparison of our DualAttNet and other SOTA attention models with FROC curve.}
  \label{fig:f5}
\end{figure}
Multi-label lesion detection in chest X-rays has received increasing attention in the field of medical image analysis. Due to the high visual similarity of various thoracic diseases in medical imaging, attention mechanism has been widely inserted into general detection architectures to highlight different lesion regions in the feature map. For existing popular methods, such operations usually bring additional time costs in the inference phase. In contrast, DualAttNet focuses on current pyramid feature-based detectors and constructs an attention sub-network and leverage loss function and the gradient from global and local attention branches to support the refinement of features from pyramid layers. In this case, the attention module is no longer responsible for generating the optimised feature map and will not degrade the inference speed. To illustrate the effectiveness of our approach, on the one hand, we have conducted a series of quantitative comparison experiments following COCO evaluation standards. On the other hand, FROC \cite{egan1961operating} is a crucial metric in clinical applications \cite{xie2021survey}. For each CXR image, it illustrates the average sensitivity of models based on different number of false positives per scan. As ChestX-ray8 and COVID-19 datasets contain too few samples in the test set which lack statistical significance, we combine all three datasets in this evaluation. Following the algorithm in the Camelyon16 challenge \cite{bejnordi2017diagnostic}, the result is presented in Figure \ref{fig:f5}. It can be observed that from an average of one false positive per scan, the sensitivity of DualAttNet is significantly higher than other methods. Consequently, our proposed model performs a much lower false positive rate with the same sensitivity.

\begin{figure}[!thbp]
  \centering
  % \includesvg[inkscapelatex=false]{dsh.svg}
  \includegraphics[width=1\linewidth]{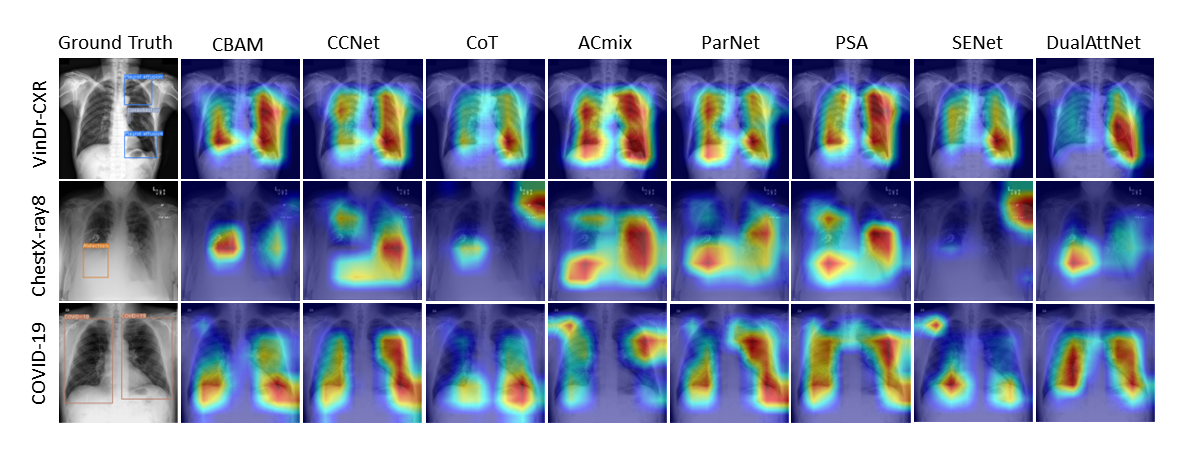}
  \caption{The Visualization of the attention regions in the method of DualAttNet vs other eight attention models.}
  \label{fig:f6}
\end{figure}
Furthermore, to further explain why the DualAttNet-based detection framework is more effective than other attention approaches in the multi-label lesion detection task, we use Eigen-CAM \cite{muhammad2020eigen} to visualise the attention region of different models on three datasets, which is shown in Figure \ref{fig:f6}. It calculates the principal components of the trained feature representations from the convolutional layers. Eigen-CAM has demonstrated robustness to misclassifications caused by fully connected layers within convolutional neural networks. Moreover, it operates independently of gradient backpropagation, class correlation scores, maximum activation positions, or any other kinds of weighted features. From the visualisation, we can see that the existing attention algorithms may not focus on correct disease areas. On the contrary, the proposed module is able to suppress other non-lesion regions and generate a more interpretable attention map. Thus, DualAttNet can reduce the detection of false positive cases and improve targeting accuracy. In addition, our proposed method still exists several limitations. Firstly, as illustrated on AP$_S$ and AR$_S$ metrics (Table \ref{tab:t2}), DualAttNet does not perform well in small object detection compared to other attention models. Secondly, although the model can be connected with most of the FPN-based detectors, some of the modern studies remove pyramid features in their architectures. Therefore, extending our method to such frameworks is also our future work. In summary, DualAttNet displays its robustness and superior performance on the multi-label lesion detection task and we believe it can be considered as a new SOTA method for computer-aid diagnosis in chest X-rays.

\section{Conclusion}
In this paper, we propose a gradient-based attention module for multi-label lesion detection in CXRs, called DualAttNet. The introduced module is comprised of an ILA block and an FGDA algorithm. The former collects global classification information from each pyramid feature layer. The latter focuses on the local attention extraction from anchor feature maps, which combines with the global attention from the ILA block, refines representations of pyramid features and guide the detector to pay attention to lesion regions using the gradient flow of binary cross entropy loss function. We evaluate our proposed method on three CXR datasets. The results demonstrate that compared to other attention models, DualAttNet not only achieves higher scores in standard mAP and AP$_{50}$ metrics but also performs lower false positive rates with the same sensitivity. In the future, we will explore the application of DualAttNet in non-FPN detection architectures and other medical image analysis tasks, such as instance segmentation.

%% Loading bibliography style file
\printcredits
\nocite{*} 
\bibliographystyle{model1-num-names}
% \bibliographystyle{cas-model2-names}

% Loading bibliography database
\bibliography{cas-refs}

\end{document}